\documentclass[journal]{IEEEtran}
\usepackage{epsfig}
\usepackage{latexsym}
\usepackage{amssymb}

\begin{document}
 
\title{On the first Townsend coefficient \\ at high electric field}
\author{Yu.I.~Davydov \\
Joint Institute for Nuclear Research, 141980, Dubna, Russia
\thanks{}
\thanks{* The study  have in the  most been  done during  author's stay  at
TRIUMF, Vancouver, BC, Canada.}}
\maketitle

\begin{abstract}

Based  on  the simplified  approach  it  is  shown and  experimentally
confirmed that gas gain in  wire chambers at very low pressure becomes
higher on thicker  wires at the same applied high  voltage.  This is a
consequence of  the fact that  the first Townsend coefficient  at high
reduced electric field  depends almost entirely on the  mean free path
of the electrons.

\end{abstract}

\section{Introduction}

An electron drifting between two  points ${r_1}$ and ${r_2}$ under the
influence  of an electric  field gains  energy and  produces secondary
electrons  due to  inelastic  collisions. The  energy distribution  of
electrons changes in  shape from the Maxwellian (at  the absence of an
applied field) to a wider distribution under the electric field.  When
attachment, photoproduction  and space charge  effects are negligible,
the multiplication factor {\it M} over that path is expressed by

\begin{equation}\label{eqn:mult1}
 lnM=\int_{r_1}^{r_2} \alpha (r)dr ,
\end{equation}

\noindent  where $\alpha(r)$  is the  first Townsend  coefficient. The
Townsend coefficient is a  function of reduced electric field strength
{\it S=  E/p}, i.e.  $\alpha  /P=f(E/p)$.  There are several  forms of
the  first  Townsend   coefficient~\cite{Town_forms}.   Most  of  them
satisfactorily  describe  experimental  data  on some  ranges  of  the
electric field  with correctly determined  parameters. The generalized
form  of  the reduced  first  Townsend  coefficient  is given  by  the
expression

\begin{equation}\label{eqn:town1}
 \frac{\alpha}{p} = A exp(\frac{-Bp}{E}) ,
\end{equation}

\noindent where  {\it A} and {\it  B} are parameters  depending on the
gas type and electric field range.

Let $\lambda$ represent an electron  free path, i.e.  the path between
two consecutive collisions with  atoms, $\lambda_E$ is a projection of
the free path on the field  direction, $\lambda_m$ is a mean free path
and $\lambda_i$ is an ionization path, i.e.  the distance in the field
direction required  for the electron to travel  between two successive
ionizations.   It  is  obvious  that  for  each  gas  $\lambda_m$  and
$\lambda_i$  depend on  pressure.  $\lambda_i$  is a  function  of the
local electric field as well.

In gases,  distances between  two consecutive collisions  of electrons
and atoms have an exponential  distribution. In general, the mean free
path in gases is defined as~\cite{Engel}

\begin{equation}\label{eqn:meanfp}
\lambda_m = \frac{1}{n\sigma },
\end{equation}

\noindent where  {\it n} is  the number of  atoms per unit  volume and
$\sigma$ is the total cross section for electron collision with atoms.
Generally,  the  cross section  is  a  function  of electron  velocity
(energy) $\sigma=\sigma(v)$.

It  has been shown~\cite{Zastawny}  that the  generalized form  of the
first  Townsend  coefficient expressed  by  Eq.(\ref{eqn:town1}) is  a
satisfactory  description of  experimental  data in  a  wide range  of
electric field.  This  form of the first Townsend  coefficient will be
used hereafter as well.

\section{Townsend coefficient at high reduced electric field}

\subsection{Simplified case}

Consider a  hypothetical simplified case  when free paths  between two
collisions of  electrons with atoms  are constant at a  given pressure
and  the probability  for  the  electron with  energy  above the  atom
ionization level to produce an  ionization at each collision is 1. Let
all electrons  have an energy  equal to the  average energy of  a real
energy  ditribution.  All  quantitative estimations  included  in this
subsection  are  made under  these  simplified  assumptions.  In  this
simplified case the free paths are equal to $\lambda = \lambda_m = 1/n
\sigma$.
 
Consider  a hypotethical  electric  field when  the  field strength  E
gradually increases from 0 to  some value. Let electrons at each point
be in  equilibrium with  an electric field.   The higher  the electric
field,  the  higher the  average  electron  energy.   At low  energies
electrons scatter  isotropically.  However,  at a high  electric field
electrons  gain  large  energy  and  already  at  energy  of  20-30~eV
scattering  takes place  mostly  in forward  direction~\cite{Vuskovic,
Boesten, Sakae}. As  a result, $\lambda_E$ tends to  $\lambda_m$ .  In
this case  at each collision electron  loses energy equial  to the gas
ionization potential.   When distances  between two collisions  in the
field direction become greater than the electron path required to gain
enough  energy  to ionize  atoms,  i.e.   $\lambda_E \geq  \lambda_i$,
electrons  restore their energy  and the  average energy  continues to
increase.   At  this  conditions  the first  Townsend  coefficient  is
expressed very simply as

\begin{equation}\label{eqn:town2}
\alpha = \frac{1}{\lambda}
\end{equation}

Therefore  the  first  Townsend  coefficient  can be  expressed  as  a
combination of two components:

\begin{eqnarray}\label{eqn:town3}
\alpha_1 &  = & Ap  exp(\frac{-Bp}{E}), \lambda_E<\lambda_i\nonumber\\
\alpha_2 & = & 1/\lambda, \lambda_E \geq \lambda_i
\end{eqnarray}

\noindent   with   the    general   gas   multiplication   factor   of
Eq.(\ref{eqn:mult1}) expressed now as

\begin{equation}\label{eqn:mult2}
 ln M=\int_{r_1}^{r_2} (\alpha_1 (r) + \alpha_2 (r)) dr.
\end{equation}

The  first term  in Eq.(\ref{eqn:town3})  is a  standard form  for the
first Townsend coefficient and describes events when there are several
collisions between  two consecutive ionizations. The  second one takes
into account  cases when there  are no elastic collisions  between two
consecutive ionizations.

An electron drifting  at the electric field {\it  E} over the distance
$\lambda_E$ gains  energy ${\it eE\lambda_E}$, where  $\lambda_E$ is a
projection of  $\lambda_m$ onn the field direction.   As was mentioned
above,  at the  high electric  field  electrons have  a large  average
energy and  scatter mostly in  the forward direction,  and $\lambda_E$
tends to $\lambda_m$.  At these conditions for any gas at a high field
there exists  an electric field  strength ${E_m}$ when  electron gains
enough energy to ionize the gas over the path $\lambda_m$, i.e.

\begin{equation}\label{eqn:em}
eE_m\lambda_m=I_0,
\end{equation} 

\noindent  where ${I_0}$ is  the gas  ionization potential.   For this
case Eq.(\ref{eqn:town3}) could be re-written as

\begin{eqnarray}\label{eqn:town5}
\alpha_1 & = & Ap  exp(\frac{-Bp}{E}) , E< E_m\nonumber\\ \alpha_2 & =
& n\sigma, E \geq E_m
\end{eqnarray}

There are very important consequences from the last equations both for
parallel plate avalanche counters (PPAC) and wire chambers.

In  case  of  the  PPAC  the  electric field  is  homogenous  and  the
multiplication  factor over the  cathode-anode distance  is expressed,
according to Eq.(\ref{eqn:town5}), as

\begin{equation}\label{eqn:mult3}
lnM = \int_{a}^{c} (Ap exp(\frac{-Bp}{E}) dr
\end{equation}

\noindent when ${\it E < E_m}$, or

\begin{equation}\label{eqn:mult4}
lnM = \int_{a}^{c} (n\sigma) dr
\end{equation}

\noindent when ${\it E \geq E_m}$.

This means  that in  the simplified case  of constant free  paths when
$\lambda  \equiv \lambda_m$  there is  a limit  of the  electric field
strength ${\it E_m}$ where  further voltage increase does not increase
gas multiplication.   This field  $E_m$ is given  by Eq.(\ref{eqn:em})
with $\lambda_m$ defined by Eq.(\ref{eqn:meanfp}).

At standard conditions for an ideal  gas the number of atoms in a unit
volume   is  $n=2.687\cdot   10^{19}$~cm$^{-3}$   (Loschmidt  number).
Generally, the cross section is  a function of electron energy and the
typical value is  of the order of 10$^{-15}$~cm$^{2}$~\cite{Palladino,
Schultz, Biagi_1}.  Typical  ionization potentials for different gases
are in the range 10-15~eV.   Substituting values for $\sigma$, {\it n}
and  $I_0$  into  Eq.(\ref{eqn:em}) gives  $E_m  \simeq$270-400~kV/cm,
which  is  far beyond  a  reachable  value  for PPACs.   However,  the
situation changes at  low pressure.  At a gas  pressure of 20~Torr, $n
\simeq $ 7 $\cdot$ 10$^{17}$~cm$^{-3}$, and Eq.(\ref{eqn:em}) gives an
electric  field  value in  the  range  $E_m  \simeq $7-10.5~kV/cm,  or
$S_m=E_c/p  \simeq   $350-520~V/cm$\cdot$Torr  in  terms   of  reduced
electric field.  These values are in the range where PPACs are used at
low pressure~\cite{Sernicki}.

More    important    consequences    from   the    consideration    of
Eq.(\ref{eqn:town5}) appear for wire chambers. The electric field in a
cylindrical wire chamber is defined as

\begin{equation}\label{eqn:efield}
E=\frac{V}{r ln(b/a)},
\end{equation}

\noindent where  {\it a} and {\it  b} are the wire  and cathode radii,
and {\it r} is the distance  from the wire center.  The electric field
strength has a maximum value on the wire surface and sharply drops off
away from the wire.  As a  result, at the atmospheric pressure the gas
gain mainly takes place within 3-5 wire radii.

Electrons  drifting toward the  wire gradually  gain energy.   For the
simplified case  mentioned above, each collision of  the electron with
energy  above the  ionization potential  results in  ionization.  When
electrons  gain energy  about few  tens of  eV, the  scattering mostly
takes place in the forward direction.

Real gas amplification starts  when the electric field becomes greater
than   some  critical  value   $E_c$.   This   critical  value   is  a
characteristic  of  the  gas  and  for  most gases  is  in  the  range
30-70~kV/cm     at    the     atmospheric     pressure,    or     $S_c
\simeq$40-90~V/cm$\cdot$Torr in terms  of reduced electric field.  The
reduced critical electric field $S_c=E_c/p$  is a constant for a given
gas and independent of pressure.

In  wire chambers  with the  same  geometry and  different anode  wire
diameters at  the atmospheric pressure  one needs to apply  much lower
voltage to chambers with thin wires in order to get the same gas gain.
At lower pressure the voltage  difference required to get the same gas
gain on two different anode wire diameters becomes smaller.

Now recall that there exists an electric field $E_m$ where an electron
gains  energy  equal  to  the  gas ionization  potential  between  two
consecutive  collisions.    Although  the  field  near   the  wire  is
non-uniform, for simplicity one can consider an average electric field
over the path  between two consecutive collisions.  We  found that for
gases  at  standard  conditions  this  field was  in  the  range  $E_m
\simeq$270-400~kV/cm or $S_m  \simeq $350-520~V/cm$\cdot$Torr in terms
of reduced  electric field.  Like the critical  reduced electric field
strength $S_c$, $S_m$ is a constant for a given gas and is independent
of pressure.

\begin{figure}
\begin{center}
\epsfig{file=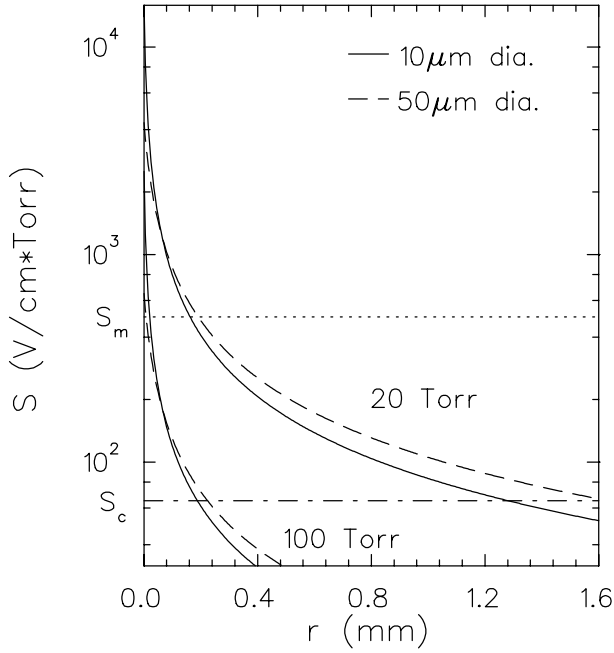,width=8cm,angle=0}
\end{center}
\caption{Reduced electric field strength as a function of the distance
from the wire surface for  single wire chambers with cathode diameters
10~mm and wire diameters 10 and 50~$\mu$m. Both wires are at 1200~V at
100~Torr and at 900~V at 20~Torr.}
\label{efield1}
\end{figure}

Consider as an example two single wire chambers with cathode diameters
10~mm     and    wire     diameters    10~$\mu$m     and    50~$\mu$m.
Figure~\ref{efield1}  presents the reduced  electric field  value near
the 10~$\mu$m and  50~$\mu$m wires as a function  of the distance from
the  wire surface.   1200~V  applied  to both  wires  at the  pressure
100~Torr  and   900~V  at  20~Torr.    We  will  use   typical  values
$S_c$=65~V/cm$\cdot$Torr  and   $S_m$=500~V/cm$\cdot$Torr  in  further
consideration. These values $S_c$ and $S_m$ are shown in the figure as
well.  For each wire let ${\it  r_c}$ be the point where $E=E_c$, i.e.
the point where the gas avalanche starts, and ${\it r_m}$ be the point
where  $E=E_m$.   Figure~\ref{efield1}  shows  that  at  100~Torr  the
avalanche starts about 0.2~mm away from the 50~$\mu$m wire surface and
about 0.18~mm away  from the 10~$\mu$m wire.  At  20~Torr these values
are 1.6~mm and 1.3~mm respectively.

\begin{figure}
\begin{center}
\epsfig{file=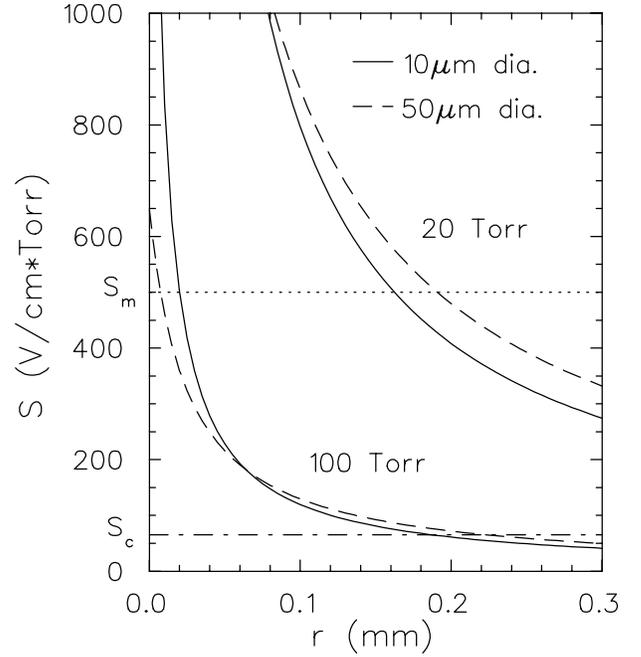,width=8cm,angle=0}
\end{center}
\caption{Details   of  the  reduced   electric  field   strength  from
Fig.\ref{efield1}  within  0.3~mm from  the  wire  surfaces. Note  the
linear vertical  scale.  Both wires are  at 1200~V at  100~Torr and at
900~V at 20~Torr.}
\label{efield2}
\end{figure}

With the same chamber geometry and applied voltages the electric field
strength is much higher on the  surface of the thin wire. It drops off
faster  on the  thin wire  and eventually  the field  strength becomes
higher  near the  thick  wire  as can  be  seen in  Fig.\ref{efield1}.
Figure~\ref{efield2} shows details of  the same field strengths with a
linear vertical scale.

The  gas gain  on  each  wire is  defined  by Eq.(\ref{eqn:mult1})  by
integrating the  first Townsend  coefficient from the  avalanche start
point ${\it r_c}$ to the  wire surface ${\it r_a}$.  In the simplified
case one can  divide this integral into two  parts.  One part includes
the path from ${\it r_c}$ to ${\it r_m}$ where $S<S_m$, and the second
one is from ${\it r_m}$ to the wire surface ${\it r_a}$. Thus, the gas
gain is defined as

\begin{equation}\label{eqn:mult5}
 lnM=\int_{r_m}^{r_c}  \alpha_1   (r)dr  +  \int_{r_a}^{r_m}  \alpha_2
 (r)dr,
\end{equation}

\noindent  where  $\alpha_1(r)$   and  $\alpha_2(r)$  are  defined  by
Eq.(\ref{eqn:town5}).

For the simplified  case at a reduced electric  field strength $S>S_m$
any electron  collision with the atom results  in ionization. However,
due to  mainly forward scattering, in  the vicinity of  a wire surface
the electrons  experience very few collisions with  atoms.  The number
of collisions depends  on the free path and in  the simplified case is
independent  of   the  electric   field,  i.e.   the   first  Townsend
coefficient becomes independent of the electric field strength.

Let us consider  in detail the electric field  strengths at a pressure
of  100~Torr and  applied  voltage  of 1200~V.   The  gas gain  starts
earlier on  the 50~$\mu$m wire. However,  the field is  very weak here
and the  contribution to the total gain  is insignificant. Eventually,
the electric  field strength  becomes higher on  the 10~$\mu  m$ wire.
Everywhere  after   the  field  crossing  point   the  first  Townsend
coefficient for the 10~$\mu $m wire is higher than or equal (after the
electric field strength on the 50~$\mu$m wire exceeds the $E_m$ value)
to  that on the  50~$\mu$m wire.   As a  result, the  gas gain  on the
10~$\mu $m wire is higher than on the 50~$\mu$m wire at this pressure.

At  a pressure  of  20~Torr and  900~V  applied to  both chambers  the
situation  is different.   As usual,  at distances  far from  the wire
surfaces the electric field strength  is higher for the thick wire and
gas gain  starts earlier  on that wire.   However, in contrast  to the
previous case the electric field strength lines cross above the $S_m$=
500~V/cm$\cdot$Torr  line.    This  means  that   the  first  Townsend
coefficient  is higher  on  the 50~$\mu$m  wire  everywhere until  the
electric  field strength  near the  10~$\mu$m wire  reaches  the value
$S=S_m$=500~V/cm$\cdot$Torr.   After that  point both  wires  have the
same  first Townsend  coefficients.  As  a  result, at  a pressure  of
20~Torr and  applied high voltage of  900~V the total gas  gain on the
50~$\mu$m wire is higher than that on the 10~$\mu$m wire.

At the atmospheric pressure an amplification in the vicinity of a wire
surface, typically within 3-5 wire  radii, is the main contribution to
the gas gain.  The higher  electric field near the thiner wire surface
results in  a higher gas gain.   At a very low  pressure the situation
changes. The thin wire has almost the same gas gain in the vicinity of
the  wire surface as  the thick  wire with  the same  applied voltage.
Differences in  the total  gas gain between  thin and thick  wires are
defined in most by the gain far from the wire surface, where the thick
wire has a higher electric field and first ionization coefficient.

Values $S_c$ and $S_m$ are constant  for each gas and vary from gas to
gas.  As a  result, different gases will have  different pressures and
applied high  voltages at which gas  gain becomes higher  on the thick
wire for the same chamber geometry and applied voltages.

\subsection{Real gas case}

In  the previous  subsection we  considered the  simplified  case with
constant electron free path lengths and electron energies equal to the
average value of the real energy distribution.  In real gases electron
free  paths  have  an  exponential  distribution  with  a  mean  value
expressed by Eq.(\ref{eqn:meanfp}).  The electrons energy distribution
is very  wide with  a long tail.   Also, the ionization  cross section
$\sigma_i$ is only  a fraction of the total  cross section $\sigma$ at
electron  energies above  the  ionization level.   At  a high  reduced
electric field  the first  Townsend coefficient at  each point  can be
defined through the mean  free path~\cite{Engel} multiplying it by the
probability that  the path  is longer than  the local  ionization path
$\lambda_i(r)$.  Multiplying this number  by $\sigma_i / \sigma$ takes
into account  the ratio of the  ionization cross section  to the total
cross  section.   In  this  case  $\alpha_2$  in  Eq.(\ref{eqn:town3})
becomes

\begin{equation}\label{eqn:town_Engel}
 \alpha_2(r)   =  \frac{\sigma_i}{\sigma}   \frac{1}{\lambda_m}  exp(-
\frac{\lambda_i(r)}{\lambda_m}) ,
\end{equation}

The first term in Eq.(\ref{eqn:town3}) and Eq.(\ref{eqn:town5}) should
be multiplied by  the probability that path is  shorter than the local
ionization        path       $\lambda_i(r)=I_0/eE(r)$.        Finally,
Eq.(\ref{eqn:town5}) transforms to

\begin{equation}\label{eqn:town6}
\alpha(r)           =           Ap          exp(\frac{-Bp}{E})(1-exp(-
\frac{\lambda_i(r)}{\lambda_m}))+               \frac{\sigma_i}{\sigma}
\frac{1}{\lambda_m} exp(- \frac{\lambda_i(r)}{\lambda_m})
\end{equation}

\noindent or, using definitions of $\lambda_i$ and $\lambda_m$

\begin{equation}\label{eqn:town7}
\alpha(r)   =   Ap    exp(\frac{-Bp}{E(r)})(1-exp(-   \frac{   I_0   n
\sigma}{eE(r)}))+ n \sigma_i exp(- \frac{I_0 n \sigma}{eE(r)})
\end{equation}
 
The first  term in Eq.(\ref{eqn:town7}) will dominate  at a relatively
low  reduced electric  field.  This  term vanishes  at a  high reduced
electric  field  and the  Townsend  coefficient  will almost  entirely
depend on the second term.

The reduced first  Townsend coefficient defined by a  standard form of
Eq.(\ref{eqn:town1})  is normally  parametrized  for some  range of  a
reduced electric field  {\it E/p}.  It tends to  a constant value {\it
A} with further increase of {\it E/p}.

The  first  Townsend  coefficient  expressed  by  Eq.(\ref{eqn:town7})
coincides  with  that expressed  by  Eq.(\ref{eqn:town1})  at a  lower
electric field. It  reaches the peak value as  {\it E/p} increases and
drops with a further increase  of the electric field, when the average
electron energy  increases.  This happens  because the second  term in
Eq.(\ref{eqn:town7})  drops due  to decrease  of the  ionization cross
section $\sigma_i$ with increasing average electron energy.

Experimental results~\cite{Beingessner} and  the numerical solution of
the     Boltzmann     equation      along     with     Monte     Carlo
calculations~\cite{Segur_1}  demonstrated   that  the  first  Townsend
coefficient {\it $\alpha$ /p} is not a function of {\it E/p} only, but
a   gradient   of   an   electric   field   as   well.    Authors   of
Ref.~\cite{Segur_1}  showed  that  at   low  pressure  there  are  two
processes which  affect the first Townsend coefficient  behaviour at a
high electric  field.  First, at high {\it  E/p} electrons approaching
the  wire could  miss  the wire  surface  and rotate  around it.   The
probability  for this  is  higher on  a  thinner wire,  which, as  the
authors noted, results  in a higher first Townsend  coefficient on the
thinner wire  at the same value  of {\it E/p}.  However,  in this case
one can talk about increase of an effective electron drift path rather
than increase of the first  Townsend coefficient.  Such an increase of
the electron  path due to rotation  around the wire  will increase the
total gas gain. Another process which results in decrease of the first
Townsend    coefficient   is    due   to    the    so-called   delayed
electrons~\cite{Segur_1}.  In a high electric field gradient electrons
are not  in equilibrium  with the electric  field and their  energy is
lower  than that  in the  same constant  field. This  results,  as the
authors  noticed,  in  decrease  of  {\it $\alpha$  /p}  with  further
increase of the electric field.

The authors  of Ref.~\cite{Pruchova} made a Monte  Carlo simulation of
electron avalanches in  proportional counters.  They demonstrated that
rotation of electrons  around the wire did increase  the gas gain, but
made a  small contribution to the  total gain.  It was  shown that the
main gain increase  happened due to extension of  the avalanche region
further from  the wire.  These conclusions are  generally in agreement
with the results shown in the present paper for the simplified case.

Microscopic calculation of the first Townsend coefficient and gas gain
as a function  of pressure and a wire diameter made  by the authors of
Ref.~\cite{Segur_2}  demonstrated  that  decrease  of a  gas  pressure
eventually resulted in  a higher gain on the thicker  wire at the same
applied voltage.

A  set  of measurements  of  gas gain  in  single  wire chambers  with
12$\cdot$12~mm cell cross sections  with different wire diameters (15,
25, 50 and  100~$\mu$m) have been made in order to  check the gas gain
behaviour at a  high reduced electric field {\it  E/p}.  Chambers were
filled with pure \mbox{iso-C$_{4}$H$_{10}$} at pressures 92, 52, 32 or
12~Torr and  irradiated with $^{55}$Fe  x-ray particles.  Most  of the
electrons released  by the photoabsorption  process in the  gas volume
leave the  chamber cell.  A small  fraction of them  lose their entire
energy inside  the cell and give full  signals.  These photoabsorption
peaks were  used to  calculate the gas  gain.  The intensity  of these
full photoabsorption  peaks drops with decreasing gas  pressure due to
the  increasing  electron  range  in  the gas.   The  results  clearly
demonstrated that a decrease of a  gas pressure from 92 to 12~Torr led
to a  higher gas  gain on  thicker wires compared  to that  on thinner
ones.    Detailed  results   of  these   measurements   are  published
separately~\cite{Davydov}.

\begin{figure}
\begin{center}
\epsfig{file=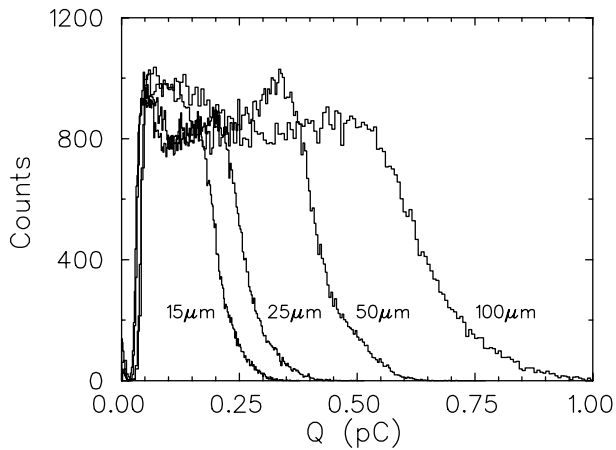,width=8cm,angle=0}
\end{center}
\caption{Measured charge spectra from  single wire chambers with cells
12$\cdot$12~mm in  size and anode wires  15, 25, 50  and 100~$\mu$m in
diameter, filled with pure \mbox{iso-C$_{4}$H$_{10}$} at 12~Torr.  All
chambers are at 800~V and irradiated with $^{55}$Fe x-rays. The larger
the wire diameter, the higher the gas gain.}
\label{12torr}
\end{figure}

Figure~\ref{12torr}  presents  one  result  from  these  measurements,
namely  the  charge spectra  taken  from  single  wire chambers  at  a
pressure of  12~Torr, with 800~V  applied to all  chambers.  Electrons
released by  the photoabsorption of  $^{55}$Fe x-ray particles  have a
range of  about 45~mm  in pure \mbox{iso-C$_{4}$H$_{10}$}  at 12~Torr.
Almost all of  them leave the chamber cell before  losing all of their
energy  and  there  is  no  evidence of  photoabsorption  peaks.   The
resulting   charge   spectra  from   the   chambers  have   continuous
distributions. However, all chambers  have the same initial ionization
distribution and the edges of the  spectra do indicate the gas gain on
each   wire.    The  figure   clearly   demonstrates   that  in   pure
\mbox{iso-C$_{4}$H$_{10}$}  at 12~Torr and  applied 800~V  the thicker
wires have a higher gas gain.   Taking a collected charge on each wire
at the  half maximum on the edge  of the spectra as  a reference gives
the     ratio    of     gas     gains    on     all    these     wires
$M_{15}:M_{25}:M_{50}:M_{100} \simeq 1:1.35:1.85:3.15$.

It   should   be   noted   that   the   chamber   simulation   program
Garfield~\cite{Garfield}  with  the Magboltz~\cite{Biagi_2}  interface
shows a  similar tendency on the  gas gain at  low pressure.  Although
Garfield  overestimates  the  first  Townsend coefficient  and,  as  a
result, the gas gain, one can compare the gas gain ratios for wires of
different diameter.  The Garfield estimation of avalanche sizes due to
single electrons in  the same wire chambers under  the same conditions
as  in   the  above-mentioned  example   gives  the  gas   gain  ratio
$M_{15}:M_{25}:M_{50}:M_{100} \simeq 1:2.18:4.79:6.47$.

\section{Conclusion}

At a high reduced electric  field, where drifting electrons have large
energy and scatter mostly in the forward direction, the first Townsend
coefficient should  almost entirely depend  on the electron  mean free
path.   This reflects the  well-known fact  that the  function $\alpha
/P=f(E/P)$  is  saturated  at  a  high reduced  electric  field.   The
generalized  formula for  the  first Townsend  coefficient  at a  high
reduced  electric  field  should  include  the measured  gas  data  as
parameters.

It is  experimentally demonstrated that  gas gain in wire  chambers at
very low pressure becomes higher  on thicker wires at the same applied
high voltage.  Thinner wires have  a much higher electric field in the
vicinity  of  the wire  surface  at  the  same applied  high  voltage.
However, the first Townsend coefficient  remains almost the same as in
the vicinity of the thicker  wires.  Very few ionizations occur in the
vicinity of the wire surface and  the total gas gain is defined mostly
far from  the wire surface where  the electric field  strength and the
first Townsend  coefficient are higher for the  thicker wire.  Thicker
wires should be  used in wire chambers operating  at very low pressure
where scattering on the wires is not critical.

In PPACs at  some value of the reduced electric  field strength in the
simplified case of constant free paths  there is a limit for gas gain,
which is  defined by gas  density, i.e.  electron's free  paths. These
values vary  from gas  to gas  and are in  the range  $S_m \simeq$340-
500~V/cm$\cdot$Torr.   In real  gases free  paths have  an exponential
distribution, and there will be  no sharp transition when the electric
field reaches the value  $S_m$.  It should asymptotically approach its
limit instead. 

The author  wishes to thank R.~Openshaw and  V.~Selivanov for fruitful
discussions and remarks.

\end{document}